# Quantum Decoding with Venn Diagrams

Caio M. F. Barros[1], F.M. de Assis[2], H.M. de Oliveira[1]

*Abstract*: The quantum error correction theory is as a rule formulated in a rather convoluted way, in comparison to classical algebraic theory. This work revisits the error correction in a noisy quantum channel so as to make it intelligible to engineers. An illustrative example is presented of a naïve perfect quantum code (Hamming-like code) with five-qubits for transmitting a single qubit of information. Also the (9,1)-Shor codes is addressed.

*Keywords— Quantum mechanics, quantum codes, stabilizer formalism.*

## I. INTRODUCTION

The theory of quantum error correction was originally discovered by Shor [1], which introduced the Shor code of nine-qubit [2] and independently by Stean [3], when studying the properties of entangled states, using different approaches. Since then, a powerful channel coding theory was born, which has been the focus of a renewed interest. The studies of this area followed, where possible, conducting approaches in standard field theory of channel coding [4-8], and thus it was possible to conceive of CSS codes (Calderbank-Shor-Stean). In the late 90's, two quantum theories, rather equivalents were presented. Gottesman [2] presented the theory of the stabilizer formalism, which was used to define code stabilizers and describe a number of quantum codes already established. Independently, a quantum theory based on properties classical geometric coding over GF(4) [9,10], allows the formulation of the first proof of a Gilbert-Varsharmov-like bound for the quantum case. Currently, the quantum theory of channel coding is moving towards its maturity. In this paper is shown the step-by-step for encoding and decoding (often omitted in quantum codes papers). The paper is organized as follows. Section II briefly reviews a number of basic principles and notation of quantum communication. Section III addresses the quantum coding and, in the next section it is shown details of a syndrome-based decoding approach. The decoding of the Shor code is presented in Section V. The final section presents concluding remarks. We adopt the symbol := to denote "equals by definition".

## II. BASIC PRINCIPLES OF THE QUANTUM COMPUTER

To insert a new paradigm in information theory – the quantum information theory – some basic concepts should be reviewed. The primary concept is the basic unit of information in this new framework, namely, the qubit. The notation |.> adopted is the so-called Dirac notation; this notation is seen often through the whole theory of quantum coding [11].

$$|\psi>:= \sum_{i=0}^{p-1} \alpha_i |i> \text{ such that } \sum_{i=1}^{p-1} |\alpha_i|^2 = 1. \quad (1)$$

As a binary digit has two possible states, 0 and 1, a qubit also has two possible states, |0> and |1>; the dissimilarity is that a qubit can be in a state known as superposition state, i.e. an intermediate state between two possible states, forming a linear combination of basis states whose coefficients are complex numbers, $\alpha_i \in \mathbf{C}$. A matrix representation for these states is: $|0>=[1\ 0]^T$ and $|1>=[0\ 1]^T$. Again, due to the similarity between bits and qubits, and the matrix representation of the basis states, we can define some quantum gates operating on a qubits-base. Equation 2 represents a quantum bit-flipping gate, and Equation 3 represents the quantum phase-inversion gate, the latter has no similarity in the classical case. But this error is as annoying as the inversion of one qubit, since the new (contaminated) state is orthogonal to the original state. There are other qubit gates, such as $Y := iXZ$, but this ensemble of quantum gates is enough in this paper. Here $i := \sqrt{-1}$.

$$X := \sigma_X = \begin{bmatrix} 0 & 1 \\ 1 & 0 \end{bmatrix} \rightarrow \begin{cases} X|0> = |1> \\ X|1> = |0>, \end{cases} \quad (2)$$

$$Z := \sigma_Z = \begin{bmatrix} 1 & 0 \\ 0 & -1 \end{bmatrix} \rightarrow \begin{cases} Z|0> = |0> \\ Z|1> = -|1>. \end{cases} \quad (3)$$

The understanding of the subject involves both the nomenclature and several physical assumptions, requiring a specialized approach, e.g. [12-13].

## III. THE QUANTUM CHANNEL CODING

The idea of quantum encoding is fairly nebulous. In this section, we propose a novel reading, more easy to understand for people who deal with the classical channel coding [14]. Consider sending a quantum information stored in a qubit that is the basis of binary states (|0⟩ and |1⟩), so the qubit to be sent is on the form $|\psi\rangle = \alpha_0|0\rangle + \alpha_1|1\rangle$. The channel in which information should be sent adds noise. In the classical case, the issue of low reliability of transmission is circumvented by introducing redundancy bits, along with a set of parity equations, so that if a mistake occurs on the sent message, errors can be detected/corrected. In the scope of quantum codes this can be accomplished by means of a remarkable property: the invariance of a quantum state over a set of evolution matrices [15]. The technique of stabilizing codes was first mentioned in the doctoral thesis of Gottesman [2].

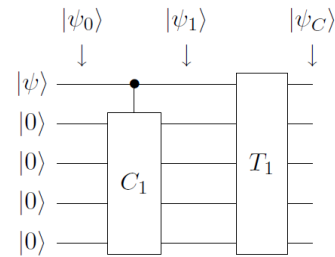

Fig. 1. Circuit for encoding the five qubits code.

It covers how one can perform an efficient coding using the invariance of the quantum state evolution matrices [11]. Along this paper we shall refer to the coding scheme for 5 qubit code (Fig. 1).

1 CMFB is with the Graduate Program in Electrical Engineering, UFPE, HmdO (Qpgom) is with DE-UFPE, Recife, PE, Brazil.
2 F.M de Assis is with IQUANTA- DEE - Federal University of Campina Grande-UFCG, PB, Brazil.

We take for granted an independent error operator such that the noise in the quantum-channel is expressed by Pauli matrices, $X$, $Z$ and $Y$. We intend to encode $k$ qubits into a new ensemble of $n$-qubits, on a non-degenerate code [2], so that each error corresponds to a subspace orthogonal to others. Thus, one can derive the following quantum bound:

$$\sum_{j=0}^{t} \binom{n}{j} 3^j 2^k \leq 2^n \text{ where } j \leq t \text{ for } t \text{ error correction.} \quad (4)$$

The bound given by Equation 4 differs from the Hamming bound [14] for block codes. Actually, there exist three distinct types of errors that can occur in a qubit; such errors are expressed by formulae 2 and 3, or a combination of these two matrices. For the case presented here, $t = 1$ and $n = 5$, indicating that it is feasible to exist a perfect code to encode one qubit to five qubit in order to correct *all* possible errors -type $X$, $Y$ or $Z$.

In the classical theory, the only codes that achieve equality in Hamming bound are those referred to as perfect codes [14]. The notation for block code is $(n, k)$, where $n$ is the blocklength of the code after the appending of parity symbols. *How many perfect quantum codes do exist*? A perfect quantum code has parameters (5,1), five qubits after encoding from one information qubit, with the ability to correct a single quantum error. There is no obvious parallel between the standard constructions of linear codes (generator matrices, polynomial generators, etc.) and the general quantum formulation. This has been a great challenge because results are not easily translated by some sort of isomorphism. The four evolution matrices of this code are presented in Table I.

TABLE I. STABILIZERS FOR THE FIVE QUBITS (5,1) QUANTUM CODE.

| Operators | Expression |
|---|---|
| $\mathcal{H}_1$ | XZZXI |
| $\mathcal{H}_2$ | IXZZX |
| $\mathcal{H}_3$ | XIXZZ |
| $\mathcal{H}_4$ | ZXIXZ |

*Information* $\psi$ (state over $k$ qubits) $\rightarrow$ *coded information* $\psi_C$ (state over $n$ qubits) $\rightarrow$ *noisy quantum channel output* $\psi_R$ (received state, possibly with error) $\rightarrow$ *measurement* of S (syndrome computing) $\rightarrow$ $\tilde{\psi}$ *decoding* (retrieved state after the decoding process).

In the case of the quantum code (5,1), an error pattern within the correction capacity of the code (that is to say, a bit inversion, a phase error from Pauli matrices $X$, $Y$ or $Z$), a decoding $\tilde{\psi}$ results exactly in the transmitted state $\psi$.

In what follows, we propose a new systematic, step-by-step interpretation, illustrating the mechanism of the quantum encoder. The first encoding step is appending four additional qubits to the information state $|\psi\rangle$, yielding $|\psi 0000\rangle$. The symbol $\otimes$ denotes the tensor product. The preparation of the state $|\psi_1\rangle = \alpha_0|00000\rangle + \alpha_1|11111\rangle$ is then carried out by means of a controlled matrix $C_1$ [11], defined by $C_1 := (X_4 X_3 X_2 X_1)^{x_5}$, Now, the coding process is performed using a coding matrix $T_1$, which is defined in terms of the stabilizers from Table I. The coded state $|\psi_C\rangle$ is therefore a superposition of two states, namely $|0_n\rangle$ and $|1_n\rangle$, which one with five qubits. The encoding follows the steps:

| | |
|---|---|
| *Information* ($k=1$) | $\|\psi\rangle = \alpha_0\|0\rangle + \alpha_1\|1\rangle$ |
| *Appending 4 qubits* ($n-k=4$) | $\|\psi\rangle \otimes \|0000\rangle = \|\psi 0000\rangle$ |
| *Controlled matrix* | $C_1 := (X_4 X_3 X_2 X_1)^{x_5}$ |
| *Applying* $C_1$ | $C_1\|\psi 0000\rangle := \|\psi_1\rangle$ |
| *Preparing the coding matrix* | $T_1 := \prod_{i=1}^{4} \left\{ \sum_{j=0}^{1} \mathcal{H}_i^j \right\}$ |
| *Coding operation* | $T_1\|\psi_1\rangle = \|\psi_C\rangle$ |
| *Coded state representation* | $\|\psi_C\rangle = \alpha_0\|0_n\rangle + \alpha_1\|1_n\rangle$. |

It is worth noting the distinction from the classical case; here the information is conveyed by a qubit, which is a state superposition through the complex numbers $\alpha_0$ and $\alpha_1$. After the encoding, states are $|0_n\rangle$ and $|1_n\rangle$, and the encoded state corresponding to a superposition of both states, where each state is encoded on a defined set of states of five qubits. This state is then encoded and may be disturbed by quantum errors.

IV. HOW TO PERFORM A QUANTUM DECODING?

How is the quantum equivalent of the parity check equations? Which operation plays the role of a parity check? The interpretation of the decoder offered here is an intuitive and more didactic interpretation. Instead of checking a parity equation (e.g.) $k_1 \oplus k_2 \oplus k_3 = c_1$, there is one between the two possibilities: the equation of stabilization [2] is checked (*pass*) or it fails (*fail*), achieving these results through all steps in the cascade with arrays of type $\mathcal{H}_i$.

> *So the proposal here is to interpret the measurements based upon the operator (stabilizers for the five qubit code, Table II), as equivalent as parity check verification.*

The decoding process begins by identifying potential errors and then correcting it, retrieving the quantum information. For the process of identification of errors we suggest to use the syndrome. The syndrome calculation is performed by exploiting the invariance property of the state encoded in relation to the matrices $\mathcal{H}_1, \ldots, \mathcal{H}_4$, Equation 5.

$$Q = \bigcap_{\mathcal{H}_i \in \Sigma} (v \in C^n | \mathcal{H}_i v = v), \quad (5)$$

where $\Sigma$ is the set of all stabilizer matrices of the code.

The measurements $\langle \psi_C | \mathcal{H}_i | \psi_C \rangle$ using the matrices $\mathcal{H}_i$, yields, if the encoded state is not modified by any error, in the scalar value 1. However, if an error corrupts the encoded state, and it presents an anti-commutative property with respect to a matrix of Table I, then the measurement using such a matrix results in the scalar -1. The syndrome is calculated in cascade by measuring the received state with the aid of the $\mathcal{H}_i$ matrices.

$$S_i = \langle \psi_R | \mathcal{H}_i | \psi_R \rangle. \quad (6)$$

Table II shows the pattern of syndromes for each type of considered error.

TABLE II. CORRECTABLE ERROR PATTERNS FOR
THE FIVE-QUBIT HAMMING-LIKE CODE.

| Errors pattern $E$ | Syndrome $S_1\ S_2\ S_3\ S_4$ |
|---|---|
| $X_1$ | 1 1 1 -1 |
| $X_2$ | -1 1 1 1 |
| $X_3$ | -1 -1 1 1 |
| $X_4$ | 1 -1 -1 1 |
| $X_5$ | 1 1 -1 -1 |
| $Z_1$ | -1 1 -1 1 |
| $Z_2$ | 1 -1 1 -1 |
| $Z_3$ | 1 1 -1 1 |
| $Z_4$ | -1 1 1 -1 |
| $Z_5$ | 1 -1 1 1 |
| $Y_1$ | -1 1 -1 -1 |
| $Y_2$ | -1 -1 1 1 |
| $Y_3$ | -1 -1 -1 1 |
| $Y_4$ | -1 -1 -1 -1 |
| $Y_5$ | 1 -1 -1 -1 |

A nice way to appreciate the mechanism of automatic forward error correction is the approach proposed by McEliece [16-19], through Venn diagrams (see also the URL http://www.systems.caltech.edu/EE/Faculty/rjm/SAMPLE_20040708.html). This process [18], applied to the (7,4) Hamming code with correction capacity $t=1$ results in 7-4=3 sets. The quantum code (5,1) results in a Venn diagram with 5-1=4 sets, as illustrated in Figure 2, inspired in McEliece representation.

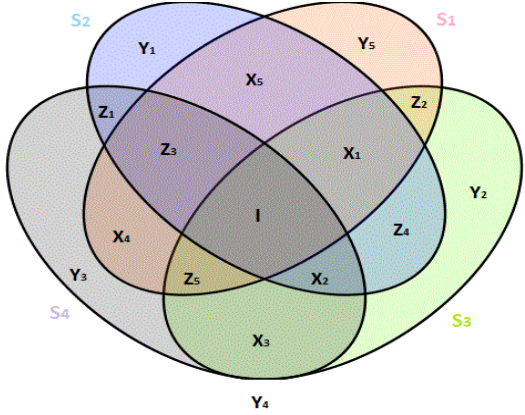

Fig. 2. Venn diagram of syndromes according with Table II for the quantum code (5,1).

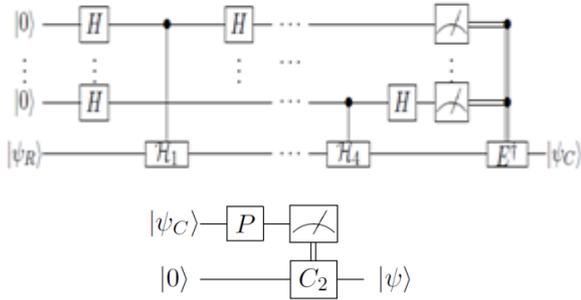

Fig. 3. a) Circuit of decoding five qubits code. b) retrieval of the transmitted qubit. H are Hadamard gates and $\mathcal{H}_i$ are stabilizer matrices.

In many published papers, there is a little obvious sense of how the decoder works. We propose an interpretation of syndrome decoding, using the standard array [7]. Once the encoded state is sent there is a probability that an error acts on it, so the received state can be modeled as $|\psi_R\rangle = E|\psi_C\rangle$. Here, as usual, [.] and {.} denote the commutative and anticommutative operators, respectively. If $[E, S_i] = 0\ \forall i$, then $E = I$; if $\{E, S_i\} = 0$ for at least some $i$, then this measure would be negative. *proof:* trivial ∎

The received state, denoted by $|\psi_R\rangle$, is obtained after passing the coded information through the quantum channel. Indeed, when there are no errors in the transmission, $E = I$ (identity) and $|\psi_R\rangle = |\psi_C\rangle$. The syndrome components are computed by measurements (Eq.6). From Table II, the error pattern $E$ that occurred in the channel is identified. Indeed, the decoder succeeds if *no errors* or *just a single* error (a pattern listed in the Table II) occurs. The error correction is carried out through the operator $E^\dagger$, defined so as to $E^\dagger E = I$.

The correction of a single quantum error corresponds to the application of the matrix $E^\dagger$ to the received state. A preparation matrix P is then defined by $P := |00000\rangle\langle 00000| + |11111\rangle\langle 11111|$ and its application to the corrected state yields an intermediate state denoted by $|\psi_2\rangle$.

A qubit is finally appended to such an state by using the tensor product $\otimes$, resulting in the auxiliary state $|\psi_{20}\rangle$. A controlled matrix $C_2 := (X_1)^{x_6 x_5 x_4 x_3 x_2}$ is then applied to the state. The concomitant collapse of the first five qubit completes the recovering of the originally transmitted quantum state $|\psi\rangle$. The sequence of operations involved in a scheme to decode the (5,1) quantum code is then summarized in the following steps:

| | |
|---|---|
| *Received state* | $\|\psi_R\rangle = E\|\psi_C\rangle.$ |
| *Syndrome computation* | $S_i = \langle\psi_C\|\mathcal{H}_i\|\psi_C\rangle$ |
| *Error identification* | $(S_1\ S_2\ S_3\ S_4) \leftrightarrow E$ |
| *Error correction* | $E^\dagger\|\psi_R\rangle = \|\tilde\psi\rangle$ |
| | $\|\tilde\psi\rangle = \|\psi_C\rangle$ if $E^\dagger E = I.$ |
| *Retrieve matrix preparation* P: | |
| *Applying* P | $P\|\tilde\psi\rangle = P\|\psi_C\rangle = \|\psi_2\rangle$ |
| *Appending* 1 *qubit* | $\|\psi_2\rangle \otimes \|0\rangle = \|\psi_{20}\rangle$ |
| *Controlled matrix* | $C_2 := (X_1)^{x_6 x_5 x_4 x_3 x_2}$ |
| *Applying* $C_2$ | $C_2\|\psi_{20}\rangle$ |
| *Collapse of the 5 qubits* | $\|\psi\rangle = \alpha_0\|0\rangle + \alpha_1\|1\rangle.$ |

## V. DECODING THE SHOR CODE

One of the most commonly cited codes, in both the classical and the quantum case is the repetition code denoted by $C(n,1)$. For instance, a $C(3,1)$-quantum code allows coding the information state $|\psi\rangle = \alpha_0|0\rangle + \alpha_1|1\rangle$ for protection against possible errors. After coding, performed by using the circuit of Fig. 4, the coded state becomes: $|\psi_C\rangle = \alpha_0|000\rangle + \alpha_1|111\rangle$

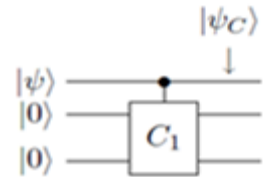

Fig. 4. Encoder for the C(3,1) quantum repetition code.

The encoding process occurs in stages. It held a coupling of two qubits in a default state, |0>, then on these coupled qubits are applied a controlled matrix $C_1=(X_2X_3)^{x_1}$ resulting therefore in the encoded state. We can see that the matrices (Table IV) when applied to the encoded state do not change it, so these arrays are termed stabilizing matrices.

TABLE IV. STABILIZING MATRICES FOR THE C(3,1) REPETITION CODE.

| Matrices | Stabilizers |
|---|---|
| $\mathcal{H}_1$ | $Z_1Z_2$ |
| $\mathcal{H}_2$ | $Z_2Z_3$ |

In order to perform efficiently the error detection it is necessary that these errors held the property of not commuting with at least one of stabilizing matrices in Table IV. An example has been the inversion of bit errors, $E_i = X_i \mid i = 1,2,3$. The Venn diagram in Figure 5 and Table V, illustrate the identification of the error depending on the syndrome circuit.

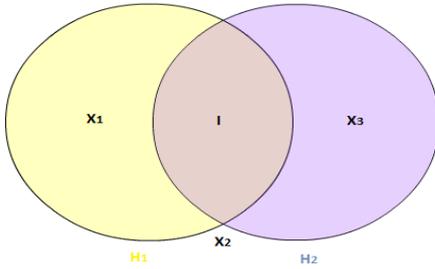

Fig. 5. Venn diagram for the C(3,1) quantum repetition code.

TABLE V. SYNDROMES FOR THE C(3,1) REPETITION CODE.

| Error pattern | $S_1S_2$ |
|---|---|
| $X_1$ | -1  1 |
| $X_2$ | -1 -1 |
| $X_3$ |  1 -1 |
| $I$ |  1  1 |

The decoding process starts with the identification of the error pattern, and after this step, the correction is performed by applying again the identified error. The corresponding decoding circuit is shown in Fig. 6.

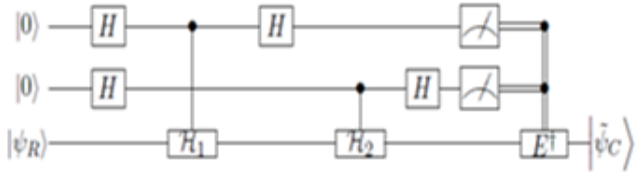

Fig. 6. Encoder Decoder for the C(3,1) repetition code.

Another case of well-established quantum code is the C(9,1) Shor code [2]. An information state $|\psi\rangle = \alpha_0|0\rangle + \alpha_1|1\rangle$ is encoded for

$$|0_L\rangle := \frac{(|000\rangle+|111\rangle)(|000\rangle+|111\rangle)(|000\rangle+|111\rangle)}{2\sqrt{2}} \text{ and} \quad (7)$$

$$|1_L\rangle := \frac{(|000\rangle-|111\rangle)(|000\rangle-|111\rangle)(|000\rangle-|111\rangle)}{2\sqrt{2}}. \quad (8)$$

This code guarantees that any types of arbitrary single error can be corrected. It is worthwhile to highlight that the standard computational basis may not be the best choice for a better understanding of the coding process so an alternative basis should be chosen. The Hadamard base is a choice, |+> and |->, where $|+\rangle := H|0\rangle$ and $|-\rangle := H|1\rangle$, where $H$ is the Hadamard matrix. Thus we can redefine the logic states of the base at the new base $|0_L\rangle := (|\tilde{+}\tilde{+}\tilde{+}\rangle)$ and $|1_L\rangle := (|\tilde{-}\tilde{-}\tilde{-}\rangle)$, states $|\tilde{+}\rangle$ and $|\tilde{-}\rangle$ are the very basis of computing the new states encoded by the C(3,1) code. The corresponding encoding circuit for the Hadamard code can be seen in Fig. 7.

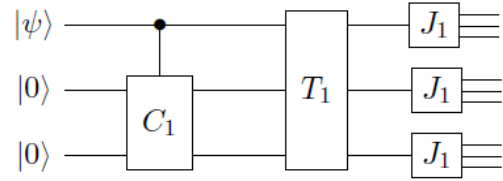

Fig. 7. Encoding circuit for the C(9,1) Shor code.

The controlled matrix $C_1$ is given by $C_1=(X_2X_3)^{x_1}$ together with $T_1=H_1H_2H_3$ change the standard basis to a Hadamard basis, and boxes $J_1$ stands for the encoder circuit of Fig.4. This process turns out the evolution matrices shown in Table VI.

TABLE VI. STABILIZER MATRICES FOR THE C(9,1) SHOR CODE.

| Matrices | Stabilizers |
|---|---|
| $\mathcal{H}_1$ | $Z_1Z_2$ |
| $\mathcal{H}_2$ | $Z_2Z_3$ |
| $\mathcal{H}_3$ | $Z_4Z_5$ |
| $\mathcal{H}_4$ | $Z_5Z_6$ |
| $\mathcal{H}_5$ | $Z_7Z_8$ |
| $\mathcal{H}_6$ | $Z_8Z_9$ |
| $\mathcal{H}_7$ | $X_1 X_2 X_3 X_4 X_5 X_6$ |
| $\mathcal{H}_8$ | $X_4 X_5 X_6 X_7 X_8 X_9$ |

Table VII expresses the syndromes for the Shor code. The syndrome calculation is carried out similarly to the previous decoding circuit and it is shown in Fig. 8.

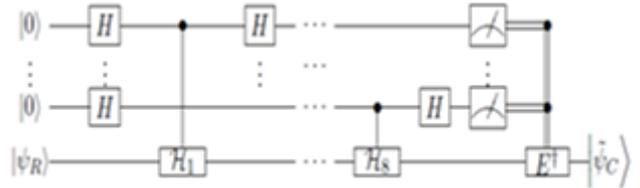

Fig. 8 Encoding circuit for the Shor C(9,1) code.

A feature that deserves a note is the fact that there exist in Table VI some rows with different error patterns, but associated with identical syndrome, e.g. the rows for $Z_1$, $Z_2$ and $Z_3$. Nevertheless, all such errors can be corrected with the same operator $Z_2$. In fact, if an error of the type $Z_1$ occurs, besides a natural correction by $Z_1$, the application of $Z_2$ yields a stabilizer (the matrix $\mathcal{H}_1$, row 1, Table VI), and the result is $\mathcal{H}_2$ when $Z_3$ occurs. A similar behavior is observed in any case for which syndromes are identical (these errors are somewhat equivalent from point of view of their effects on the coded message).

TABLE VII. SYNDROMES FOR THE C(9,1) SHOR CODE.

| Error | $S_1$ $S_2$ $S_3$ $S_4$ $S_5$ $S_6$ $S_7$ $S_8$ |
|---|---|
| $X_1$ | -1  1  1  1  1  1  1  1 |
| $X_2$ | -1 -1  1  1  1  1  1  1 |
| $X_3$ |  1 -1  1  1  1  1  1  1 |
| $X_4$ |  1  1 -1  1  1  1  1  1 |
| $X_5$ |  1  1 -1 -1  1  1  1  1 |
| $X_6$ |  1  1  1 -1  1  1  1  1 |
| $X_7$ |  1  1  1  1 -1  1  1  1 |
| $X_8$ |  1  1  1  1 -1 -1  1  1 |
| $X_9$ |  1  1  1  1  1 -1  1  1 |
| $Z_1$ $Z_2$ $Z_3$ |  1  1  1  1  1  1 -1  1 |
| $Z_4$ $Z_5$ $Z_6$ |  1  1  1  1  1  1 -1 -1 |
| $Z_7$ $Z_8$ $Z_9$ |  1  1  1  1  1  1  1 -1 |
| $I$ |  1  1  1  1  1  1  1  1 |

## VI. CONCLUSIONS

The challenge in dealing with quantum codes is often the intricacy involved. Even engineers who are rather familiar with the channel coding may find it hard to understand the principles and the mechanisms of quantum codes. Particularly, there is a little obvious sense of decoding in the many articles on the subject. This was the challenge that led the authors to make an effort to provide a formalism of quantum codes under an "understandable perspective". After all, following the traditional quote from Rutherford, we know that "*if you cannot explain your physics to the barmaid, it is probably not very good physics.*" We present here a quantum decoding with an insightful vision. The idea of the decoder is clarified based on the standard array and the syndrome calculation, reinterpreting the use of stabilizers as some sort of "parity check equations". The syndrome component calculation is the output of a quantum measurement. Despite the instance being rather naive, the results can be extrapolated to the general case. The goal is to make the mechanism much clear, with a description in language closer to the classical theory of codes, trying not to get lost in the intricacies of quantum physics. A promising extra idea is to examine how to derive a set of stabilizers from a given Venn diagram associated with a code.

ACKNOWLEDGMENTS

This study was partially supported by the Brazilian National Council for Scientific and Technological Development (CNPq) under research grant No. 305639/2009-9.